# Cross-disciplinary Reactor-to-Repository Framework for Evaluating Spent Nuclear Fuel from Advanced Reactors

Haruko M. Wainwright[1,2], Chloe Christiaen[3], Milos Atz[4], John Sebastian Tchakerian[1], Jiankai Yu[1], Gavin Keith Ridley[1], Koroush Shirvan[1]

[1] Department of Nuclear Science and Engineering, Massachusetts Institute of Technology

[2] Department of Civil and Environmental Engineering, Massachusetts Institute of Technology

[3] École Polytechnique

[4] Ultra Safe Nuclear Corporation

* Haruko M. Wainwright

Email: hmwainw@mit.edu

**Abstract**

This study presents a cross-disciplinary reactor-to-repository framework to compare different advanced reactors with respect to their spent nuclear fuel (SNF). The framework consists of (1) OpenMC for simulating neutronics, fuel depletion, and radioactive decays; (2) NWPY for computing the repository footprint for SNF disposal given the thermal constraints; and (3) PFLOTRAN for simulating radionuclide transport in the geosphere to compute the peak dose rate, which is used to quantify the repository performance and environmental impact. We first perform the meta-analysis of past comparative analyses to identify the factors led previously to inconsistent conclusions. We then demonstrate the new framework by comparing five reactor types. Significant findings are that (1) the repository footprint is neither linearly related to SNF volume nor to decay heat, due to the repository's thermal constraint, (2) fast reactors have significantly higher I-129 inventory, which is often the primarily dose contributor from repositories, and (3) the repository performance primarily depends on the waste forms. The TRISO-based reactors, in particular, have significantly higher SNF volumes, but result in smaller repository footprints and lower peak dose rates. Our analysis highlights the diversity of these reactors, each of which should be evaluated individually. The open-source framework ensures proper cross-disciplinary connections between reactor simulations and environmental assessments, as well as the transparency/traceability required for such comparative analyses. It aims to support reactor designers, repository developers and policy makers in evaluating the impact of different reactor designs, with the ultimate goal of improving the sustainability of nuclear energy systems.

## 1. Introduction

Small-modular reactors (SMR) and micro nuclear reactors (MNR) are defined as nuclear reactors ranging from 30 MWth to 1000 MWth and from 2 MWth to 40 MWth, respectively (1, 2). These advanced reactors have a significant potential to reduce carbon emissions, as well as to provide low-emission energy access to remote communities and industrial activities isolated from the main electricity grid (3). They can also be used for industrial production, water desalination, hydrogen production, and other applications (4). The leading reactor concepts include small pressurized water reactors (PWR), high-temperature gas cooled reactors (HTGR), heat-pipe reactors (HPR) and sodium-cooled fast reactors (SFR) with the fast-neutron spectrum (2). These reactors have significantly more robust safety profiles, owing not only to their small power rating, but also to their passive safety systems, high thermal capacity in the core, new fuel forms such as TRISO, and (in some cases) low-pressure operation enabled by nonvolatile coolants such as sodium (5).

As these technologies get closer to commercialization, their waste management is becoming an increasingly important topic. In 2023, the US National Academy of Science and Engineering raised concerns about waste management and disposal requirements related to SMR/MNRs (6). In particular, the report highlights their potential to increase difficulties associated with the disposal of spent nuclear fuel (SNF). Today, about 88,000 metric tons of SNF from commercial reactors are stored in dry casks and pools at reactor sites in the US. Although there is a consensus that deep geological repositories are required to isolate SNF from the environment for an extended period, there is only one country, Finland, that has successfully constructed a repository.

Several studies have investigated SNF from SMRs/MNRs, reaching divergent and contradictory conclusions. Krall et al. (7) concluded that SMRs would generate more SNF than conventional power plants and would exacerbate the challenges of nuclear waste disposal and management. Results obtained in Keto et al. (8) are in agreement with Krall et al. (7), even though their conclusion is less pessimistic. In fact, Keto et al. (8) concluded that the low discharge burnup would facilitate final disposal, since the radioactivity and the decay heat would be lower compared to conventional reactors. Brown et al. (9) also agreed with these two studies, noting that the reduction in the core size would increase neutron leakage, and hence would reduce the fuel utilization and discharge burnup. On the other hand, Kim et al. (10) concluded that SNF from SMRs would not increase challenges compared with conventional reactors. Although they agreed that small-modular PWRs (SPWR) would slightly increase the SNF volume, they found that their radiotoxicity, decay heat, and radioactivity would be comparable to those of large PWR reactors. Such inconsistent conclusions have resulted in confusion within and beyond the nuclear community. Such confusion may ultimately distort public perception regarding the environmental impacts of SMRs/MNRs, and threaten their public acceptance (e.g., 11).

The objective of this study is to develop a cross-disciplinary, open-source, reactor-to-repository framework for comparing SNF from different reactor technologies, with a particular focus on the final disposal conditions. Our framework (Figure 1) includes (a) reactor physics models and post-discharge decay calculations using OpenMC (12, 13), (b) repository footprint models using NWPY (14, 15), and (c) SNF repository performance assessment (PA) models with explicit radionuclide release/transport simulations using PFLOTRAN (16, 17, 18). All our codes are open sourced and freely available, which creates a transparent and more objective methodology for performing a comparative study.

Here our unique contributions—compared to previous studies—are the evaluation of not just the amount of waste but also the repository footprint and radiological risk associated with the SNF repository (Figure 2). As noted by Apted et al. (19), the conventional SNF metrics fail to include the repository conditions and the mobility of radionuclides, which greatly influence the size of the repository and the quantity of radionuclides that may be released to the biosphere. Even though previous studies have suggested the impact of decay heat on the repository footprint and acknowledged that radiotoxicity would not represent the environmental impact (7), none of them included repository models that would quantify those aspects.

In the following sections, we first provide a systematic review and meta-analysis of the previous studies on SMR/MNR waste (7, 8, 9, 10). Qualitative and quantitative analyses are performed to resolve contradictions among those studies, and to explain why their conclusions diverged. We then demonstrate the new framework by comparing multiple SMR/MNRs with a reference PWR based on their generic reactor designs. We consider a once-through fuel cycle, which is considered as the near-to-intermediate term choice in the US (6).

## 2. Reactor-to-Repository Framework

In our framework (Figure 1), OpenMC first simulates reactor physics, including neutronics, depletion analysis, radionuclide generation, and decay-chain calculations during the reactor operation and post-discharge periods. It provides the radionuclide composition of SNF at any given time, as well as the decay heat, activity, and radiotoxicity of each radionuclide. These can be converted to commonly used SNF metrics, including SNF mass (heavy metal equivalent) and volume as well as total activity, radiotoxicity, and decay heat (*SI Appendix* Text S1 and S2).

The repository footprint model (NWPY) then quantifies the repository area per package and per GWe.y given the thermal constraints (14, 15). It assumes a geological repository below the water table, which has been considered in Sweden, Finland, and France, as well as in the recent US generic PA models (18, 20). The repository includes engineered barrier systems (EBS), including waste forms, waste packages, and clay buffer for transport retardation. The temperature limit—for preventing the degradation of EBS—is assumed to be 100 deg-C in the clay geologic setting, which is the most stringent due to the low heat conductivity of clay.

Given the decay heat per package, NWPY solves analytical heat-conduction equations to determine the surface temperature of a waste package located at the center of a square array of *N-by-N* waste packages (*SI Appendix* Text S3: Figure 2a), where the heat contribution from adjacent packages is the highest. The package at the center is modeled as a finite line source, while the remaining $N^2-1$ packages are modeled as point sources. In addition, the drift spacing and the package spacing within the drift defines the minimum repository size. The area per package is calculated through an optimization algorithm to find the minimum area to satisfy the temperature constraint. In addition, we explore different package loading (i.e., the number of fuel elements per package) based on the previously studied package designs (*SI Appendix* Table S4; 21).

Subsequently, the generic repository PA is developed (*SI Appendix* Texts S4) to compute the release of radionuclides from the waste forms, their migration in the geosphere, and the peak dose rates over one million years (22, 23). Our conceptual model (Figure 2b) is the simplified version of the generic clay model described in Stein et al. (20), such that only the clay host rock and limestone aquifer below are considered. The release rates from the packages and repository—dependent on waste forms—are computed based on the analytical solutions developed by Ahn (21). The release rates are then connected to the groundwater flow and reactive transport code PFLOTRAN, which has been extensively used for the repository PAs (18). We focus on iodine-129 (I-129), which is typically the dominant dose contributor in the PAs (7, 18, 23, 24, 25).

The peak dose rates are then computed based on the assumption that people use groundwater pumped from a well 10 km from the repository. The groundwater concentration is converted to the dose rate using the biosphere dose conversion factor used in the US Yucca Mountain (YM) assessment (26). To compare our results with the current US regulatory standard (0.15 mSv/yr for the first 10,000 years) established for the YM repository, we multiplied the source terms to match the capacity of 70,000 metric tons (heavy metal equivalent). We would note that, for the real repository, the peak dose rates are not linear with the SNF amount, since the dose rates depend on the repository configuration and flow direction (27). I-129 is not solubility-limited, so that such multiplication can be justified. This provides the conservative prediction of the dose rates.

## 3. Meta-analysis of Previous Comparative Studies

To investigate the contradictions among the previous studies that compared the SNF from SMR/MNRs (7, 8, 9, 10), we first gather the data and assumptions from these studies, including reactor designs and their main parameters. Table 1 shows that these studies selected different reactor concepts and design parameters. In addition, the reference PWR burn-up (the key parameter determining the amount of SNF) is different among the studies, thus biasing their conclusions.

The common SNF metrics (*SI Appendix* Text S1) are then compiled and discussed in the following subsections (Table 2). We note that Brown et al. (9) and Krall et al. (7) considered thermal energy as the comparison basis (thermal efficiency was neglected), while Kim et al. (10) and Keto et al. (8) used the electricity output. We re-calculated and re-scaled these metrics based on the electricity production (GWe.year).

**Mass:** We confirmed that the SNF mass (heavy metal equivalent) is inversely proportional to the burn-up and thermal efficiency for all the reactors. Overall, the reactors with high burnup and high thermal efficiencies (Natrium and X-1000) produce lower SNF mass. Krall et al. (7) came to a pessimistic conclusion about SMRs, since their reference reactor had a relatively higher burnup, and they compared it to SMRs with relatively lower burnup. In contrast, the burn-up of 49.5 MWd/kgU for NuScale in Kim et al. (10) leads to a SNF mass similar to the reference PWR.

**Volume:** The SNF volume is impacted not only by SNF mass but also by fuel types and densities. The two fast neutron-spectrum reactors (Natrium and 4S30) have consistently small volume in Krall et al. (7) and Kim et al. (10) because of the high-density metallic fuel. For Xe-100, the high volume-to-mass ratio of TRISO pebbles increases the volume by a factor of 12.3 compared to the reference PWR.

In addition, the SNF volume is significantly impacted by the methodology used in its calculation. Krall et al. (7) considered the entire core, including the void space, while Kim et al. (10) used the assembly/pebble volume-to-mass ratio. This is why Krall et al. (7) estimated that SNF volume from NuScale was 2.5 times larger than the reference PWR, although in Kim et al. (10) the difference was only 10%.

**Decay Heat:** The decay heat at 100 years—affected by the amount of fission products and transuranic elements—is similar among the three studies for the reference PWR and NuScale. The higher burnup reactors result in increased decay heat per unit mass of initial uranium, although this figure is counterbalanced by a lower SNF mass generated per unit of electricity. There is a large discrepancy for the two fast reactors between 4S30 in Krall et al. (7) and Natrium in Kim et al. (10) such that 4S30 has 50% higher decay heat than the reference PWR, while Natrium has 50% lower decay heat. Krall et al. (7) did not perform depletion simulations but took decay heat from the MOX-based fuel (28), which has higher Pu than the U-Zr fuel that 4S30 plans to use. Kim et al. (10) noted that the lower SNF decay heat of Natrium and HTGR is attributed to the high thermal efficiency. We note that while these papers mention the impact of decay heat on the repository area, none of them had repository models.

**Activity/Radiotoxicity:** The SNF activity and radiotoxicity are governed by fission products in the first hundred years, and then afterwards by actinides and transuranic isotopes. Since NuScale and the reference are both light water reactors with thermal neutron spectrums, their isotope composition is similar. The three studies (7, 9, 10) are consistent, such that the NuScale's activity and radiotoxicity are 5–15% higher than the reference PWR. Once again, the efficiency and the discharge burnup are the main drivers of the difference. Since SFR has higher plutonium content with fast neutron spectrum, the activity and radiotoxicity are significantly higher than the reference at 100–10,000 years. In contrast, HTGR has significantly lower activity and radiotoxicity, since more Pu is consumed during the operation.

## 4. Results from the Reactor-to-Repository Framework

To demonstrate the framework, we have selected the following five relatively mature reactor concepts (Table 3) to represent the diversity of SMR/MNR designs (2, 29, 30): small PWR, HTGR, HPR, HTGR with fully ceramic microencapsulated (FCM) fuel, and SFR, compared with the reference PWR of AP1000 (31)

### 4.1. Overall Metrics

The SNF metrics are summarized in the radar chart (Figure 3).

**SNF Mass and Volume:** Compared to the reference PWR, SPWR generates 3.3 times larger SNF mass due to its lower discharge burnup. On the other hand, owing to their high thermal efficiency and/or higher burnup, HTGR, HPR, and FCM-HTGR produce 40-50% less SNF mass. In addition, with 120 MWd/kgU burnup, SFR produces the lowest SNF mass (65% lower than the reference). The SNF volumes vary more significantly. The SNF volume from HTGR, HPR, and FCM-HTGR is 25–30 times higher than the reference, due to their use of low-density TRISO fuels. In addition, SFR shows the lowest SNF volume (30% lower than the reference).

**Decay heat:** At 50 years after the discharge, the decay heat ranges from $7.7 \times 10^3$ W/GWe.y for SFR to $16.8 \times 10^3$ W/GWe.y for FCM-HTGR. This difference results primarily from the variation in the fission products, Pu-238, and Am-241. Compared to the reference PWR, SPWR produces about 15% less decay heat. As the concentration of fission products is roughly proportional to the number of fissions and burnup, the observed variation is attributed to the actinide concentrations that depend on the fuel residence time and burn-up. The decay heat is similar for HTGR and HPR, which is 15% lower than the reference due to their high efficiency. The SFR decay heat is 45% lower than the reference, due to the lower concentrations of Pu-238 and Am-241 per GWe.y. FCM-HTGR has 20% higher decay heat than the reference owing to the higher uranium density in the fuel and increased productions of actinides.

**Radiotoxicity:** At 10,000 years after the discharge, the long-term radiotoxicity varies between $8.95 \times 10^7$ Sv/GWe.y for HTGR to $4.63 \times 10^8$ Sv/GWe.y for SFR. The radiotoxicity is dominated by Pu-239 and Pu-240, accounting for more than 90% of total radiotoxicity for all the reactor types. The SPWR radiotoxicity is 60% higher compared to the reference, which is associated with the increased Pu-239 per GWe.y due to lower burn-up. The radiotoxicity for HTGR is 10% lower than the reference owing to the higher burnup and thermal efficiency. The higher radiotoxicity of SFR (360%) is attributed to the Pu-239 and Pu-240 concentration resulting from the fast neutron spectrum.

**Repository footprint:** The minimum repository footprint was defined based on the fixed interim storage time of 50 years. The footprint ranges from 757 $m^2$/GWe.y for SFR to 1472 $m^2$/GWe.y for FCM-HTGR. Compared to the reference PWR, the repository footprint is 30% smaller for HTGR, and 16% smaller for HPR. FCM-HTGR is larger than the reference by 30%. More detailed results are shown in Section 4.2.

**Peak dose rate:** The peak dose rates from the 1GWe.y-equivalent SNF range from $2.2 \times 10^{-20}$ mSv/yr for HTGR to $2.2 \times 10^{-16}$ mSv/yr for SFR. The TRISO-based reactors (HTGR, FCM-HTGR, HPR) have the values ~4 orders of magnitude lower than the others. This large difference is primarily attributed to the waste form. More detailed results are shown in Section 4.3.

**4.2. Repository footprint analysis**

The required repository area per package decreases as a function of surface storage time (Figure 4a only for HTGR as an example), because it allows the short-lived fission products to decay. As expected, the larger loading requires a larger area per package. However, the relationship is nonlinear: for the waste packages containing fewer elements and beyond a certain storage time, the decay heat is no longer constraining the spacing between the packages. The required area reaches the minimum defined by the package geometry and the drift radius.

The repository footprint is the area per package multiplied by the number of packages (Figure 4b). Since the number of packages increases with smaller loading, the 6-element loading case results in the largest footprint, which does not depend on the storage time, since the area is determined by the tunnel geometry. In contrast, with a sufficient storage time, the repository footprint decreases with increasing the number of elements. Such nonlinear relationships and the trade-off between the area per package and the number of packages lead to the minimum repository footprint. The 42-element packages result in the smallest footprint for HTGR up to 50 years of storage time. For the minimum repository footprint with a fixed interim storage time of 50 years, the optimized loading varies significantly among the reactors, from 1 to 42 (Figure 4c).

**4.3. Repository Performance Assessment**

The three waste forms—$UO_2$, TRISO, and metallic matrix—have significantly different matrix dissolution rates (Figure 5a). The metallic matrix completely disappears within 4000 years after package failure. $UO_2$ is more resistant to degradation: it takes 800,000—$8\times10^6$ years for the reference PWR and SPWR SNF to completely dissolve. TRISO performs the best, as the degradation is minimal for $10^7$ years in both the low/high degradation cases.

The cumulative I-129 release from the repository per GWe.y depends on both the initial inventory of SNF and the degradation rate (Figure 5b). The degradation rate determines the timing, while the initial inventory of I-129 determines the total release. The release from SFR SNF occurs the earliest due to its metallic waste form, and also becomes the highest due to the high inventory, since the I-129 fission yield is higher for Pu-239 (1.406%) than U-235 (0.54%). For the $UO_2$-based reactors, the release increases around 10,000 years, reaching the full release around one million years. The TRISO-based reactors (HTGR, HPR, FCM-HTGR) have small releases compared to the other waste forms.

In addition to the waste form, the geosphere provides a natural barrier against radionuclide transport. The annual dose rates associated with I-129 from using the pumped well water at 10 km from the repository start to increase after 100,000 years (Figure 5c), even for SFR with the fast-degrading metallic waste form. The dose rates at one million years vary by more than four orders of magnitude from the lowest (HTGR) to the highest (SFR). The TRISO-based reactors show the lowest concentrations compared to the other waste forms. The peak dose rates from the YM-equivalent capacity of 70,000 MTHM (metric ton of heavy metal) vary between $10^{-14}$ and $10^{-9}$ mSv/yr (Figure 5d), which are smaller than the YM standard by a significant margin.

**5. Discussion**

In this study, we first synthesized the results from the previous studies that compared the SMR/MNR SNF (7, 8, 9, 10). Our analysis showed that different reactor types and parameters were selected, even though each study made sweeping conclusions about "SMR wastes." In particular, thermal efficiency is not included in several studies, although it is an important parameter when we define the waste mass/volume per energy production. In addition, we found that there were differences in the methodology calculating the SNF volume, based on the core volume or the assembly/pebble mass-to-volume ratio. At the same time, the reference PWR was also different among the studies, which led to diverging conclusions. Finally, none of the studies included the repository models, although they implicitly related the decay heat to the footprint, and discussed the environmental impact. This is a significant omission, since the repository conditions and the mobility of radionuclides greatly influence the footprint and the quantities of radionuclides that leave the facility and enter the biosphere.

To overcome the subjectivity of previous studies and reflect realistic SNF disposal conditions, we have developed an open-source reactor-to-repository framework. It takes advantage of open-source code developments for both reactor physics (OpenMC) and repository assessments (NWPY, PFLOTRAN). Our framework takes the outputs from OpenMC and calculates various SNF metrics, and then connects the source terms to the repository footprint and radionuclide transport analyses. This framework would be useful for the industry and research communities, and valuable for educators. In particular, Wainwright et al. (32) has highlighted the need for nuclear engineering students to know more about the mobility of radionuclides and disposal conditions, and to consider the impact of reactor designs on nuclear waste for improving the sustainability of energy systems.

The SNF mass and volume results in our study are consistent with Kim et al. (10), driven largely by the burnup and thermal efficiency. The SPWR increases the SNF mass and volume compared to the reference PWR due to the lower burnup. On the other hand, the high burnup reactors have significantly lower SNF mass due to increased fuel utilization. However, the SNF volumes are significantly higher for reactors using TRISO fuel, which could lead to higher costs for handling, transportation, and storage of SNF.

Decay heat and radiotoxicity mostly depend on the fission products, plutonium and americium isotopes. Since the fission-product concentrations are nearly proportional to the energy generated, their difference is small among different reactors in the comparison. However, the thermal-spectrum reactors tend to have more Pu-238 and Am-241, which contribute significantly to the decay heat in the first 100 years (33). On the other hand, the long-term radiotoxicity is higher for SPWR and SFR, because of the higher normalized Pu-239 content in SNF. In addition, FCM-HTGR has higher radiotoxicity and decay heat, because of its higher content of Pu-238 and Am-241. This is consistent with Lu et al. (34) which has reported that the higher uranium density in the uranium nitrate fuel leads to higher Pu concentration than the $UO_2$ fuel.

One of our significant findings is to show that the repository footprint is not linearly dependent on the decay heat, since it also depends on the thermal constraints of the repository and the decay heat per package. Particularly for HTGR, although the SNF volume is significantly larger than the reference PWR, the lower decay heat density allows for tighter spacing between packages and results in a reduced repository footprint. The repository footprint is smaller for the reactors selected in this study than for the reference, except for FCM-HTGR, which has higher decay heat. HTGR indeed has the smallest repository, even though it has a 30 times larger volume than the reference.

It has been known that radiotoxicity does not represent the environmental impact or human health risk from the SNF disposal, since the majority of radionuclides are relatively immobile in the geosphere (7, 19). Although Krall et al. (7) mentioned that the mobile fission products are proportional to the number of fissions and hence to the energy generated, our analysis shows that there is a difference in the I-129 inventory between thermal and fast reactors, because of the significantly higher fission yield from Pu-239 than from U-235. In addition, our results highlight the importance of waste forms for repository performance, which is consistent with Atz et al. (35). The robustness of TRISO in capturing mobile elements and its low degradation rate (36) should considered in the SNF analysis, while the metallic fuel would require additional processing into better waste forms (37). We would note that I-129 separation and transmutation are the active area of research (38, 39)

Overall, our results have emphasized the granularity and the specificities of each reactor type, highlighting that we should avoid making sweeping conclusions about SMR/MNRs. Each reactor and its waste management should be treated and evaluated differently. In addition, we need to recognize the trade-offs involved with each rector type. The TRISO-based fuel, for example, would increase the SNF handling cost owing to the large volume, but reduce the repository footprint and improve the repository performance.

We acknowledge that our study does not consider the secondary waste streams associated with these reactors. Concerns have been raised about the new waste streams of low-level waste produced by SMR/MNRs (7). In this study, we focused on SNF as the first step, since SNF management and disposal are typically considered the most important barrier to nuclear energy expansion. In addition, uranium utilization should be included, since the front-end waste generates the largest environmental impact throughout the fuel cycle (40).

Our framework can be extended in the future to consider such secondary waste, as well as the entire fuel cycle. The reactor safety metrics—for example, the accident tolerance of TRISO—could be included to discuss the tradeoff between reactor safety and waste volume. Furthermore, the environmental impacts should be compared with other energy systems that tend to create a large amount of effluents and hazardous waste (41)—waste that has been less rigorously managed over the short compliance period of 30 years (42). Our reactor-to-repository framework aims to be a part of the effort to improve waste management and life cycle assessments, as well as to enhance the sustainability of energy systems.

**Acknowledgments**
This work was supported by MIT's internal grant. We thank Dr. Jacopo Buongiorno for productive discussions and constructive comments, Dr. Emily Stein for the support on PFLOTRAN simulations, and Dr. Daniel Hawkes for technical editing. The codes and input files are available at https://github.com/hmwainw/R2R4SNF.
**Author Contributions:** HMW and CC worked on the conceptualization, methodology and analysis as well as the writing of the original draft. MA developed the repository footprint model, and did the reviewing/editing of the draft. KS designed the reactor physics models as well as did the reviewing/editing of the draft. JST, JY, and GK revised the reactor simulation input files, and ran simulations with CC as well as did the reviewing/editing of the draft.

**Figures and Tables**

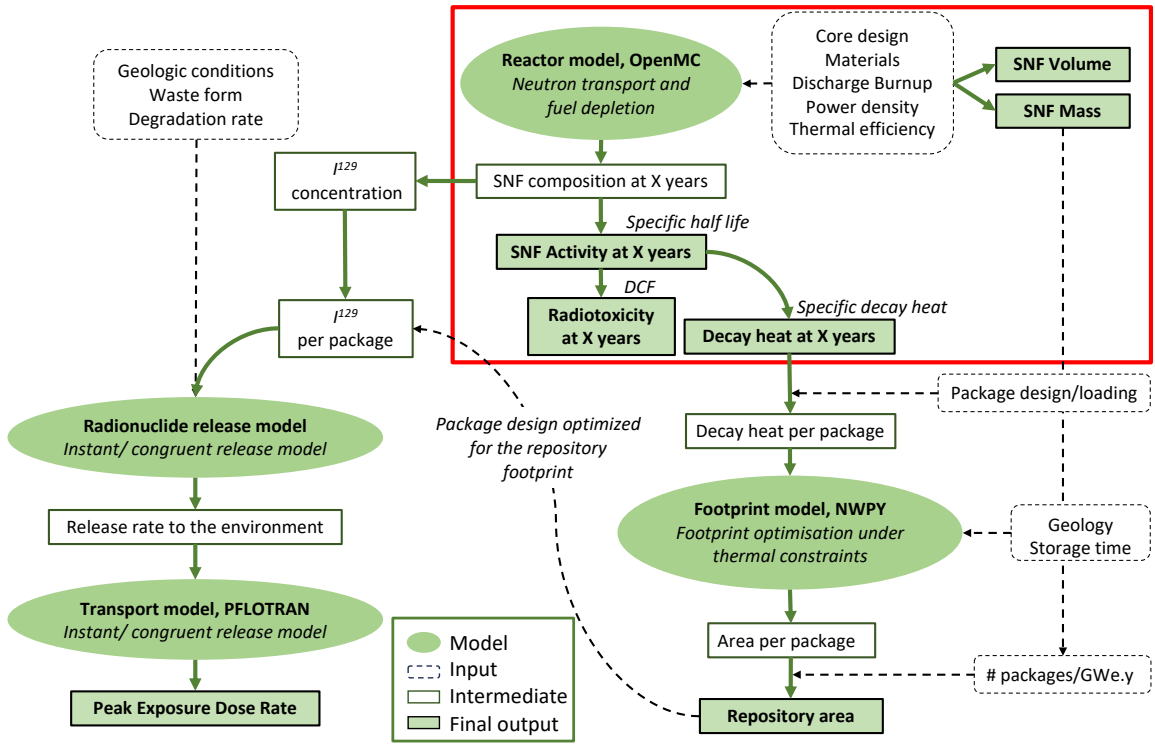

**Figure 1.** Diagram summarizing the various inputs, steps, and models used in our framework. The red box represents the components typically included in the previous comparative analysis of SMR/MNR SNF; the rest is newly developed in this study. DCF is the dose conversion factor.

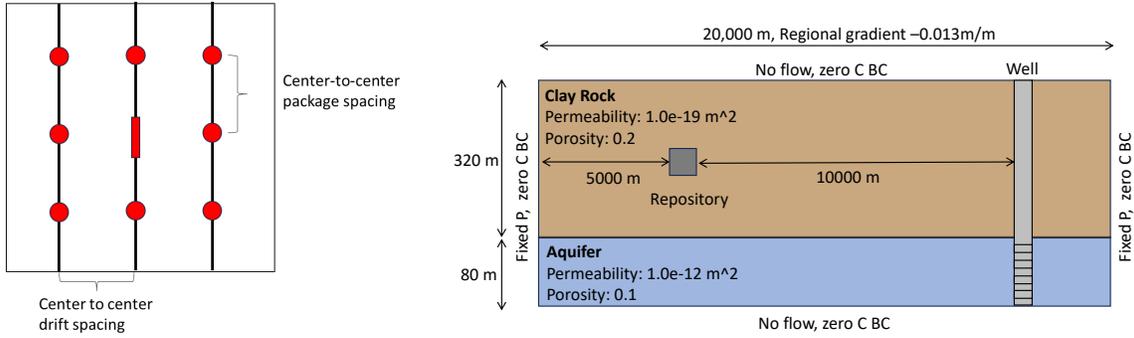

**(a)**                          **(b)**

**Figure 2.** (a) A repository footprint model with the layout of a 3x3 repository package array, and (b) a simplified performance assessment (i.e., radionuclide transport) model from a generic saturated-zone repository. In (a), the central package is modeled as a finite line source, while adjacent canisters and drifts are modeled as point sources. In (b), the repository parameters are from Stein et al. (20).

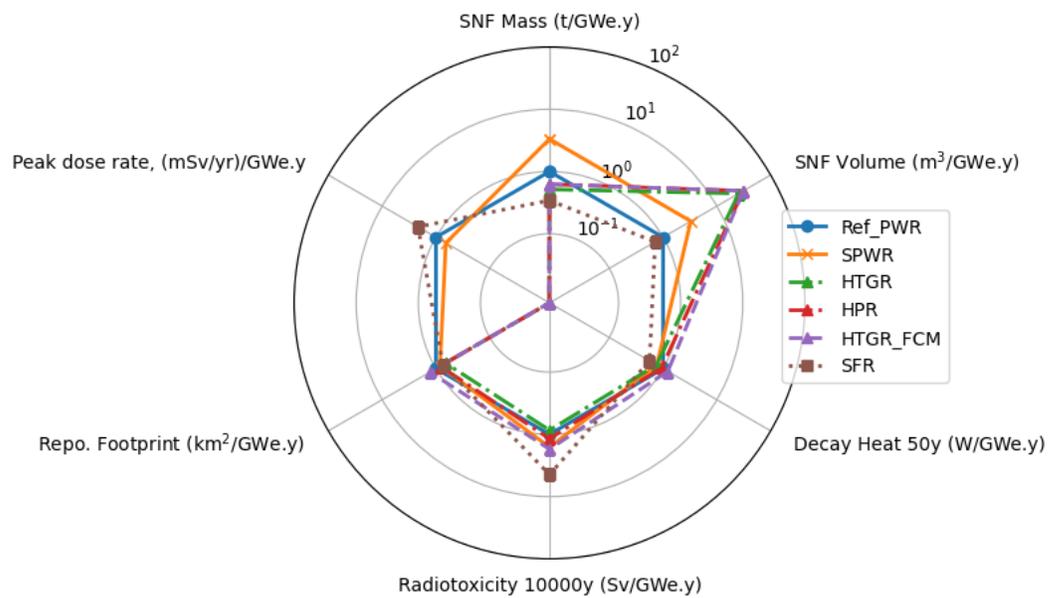

**Figure 3.** Radar chart comparing SNF from different reactor types normalized against the respective results for the large reference PWR reactor. The units shown are the computed values before normalization. The reactors with the same type of fuel are represented with the same line styles. Similar burnups are showed with the same markers.

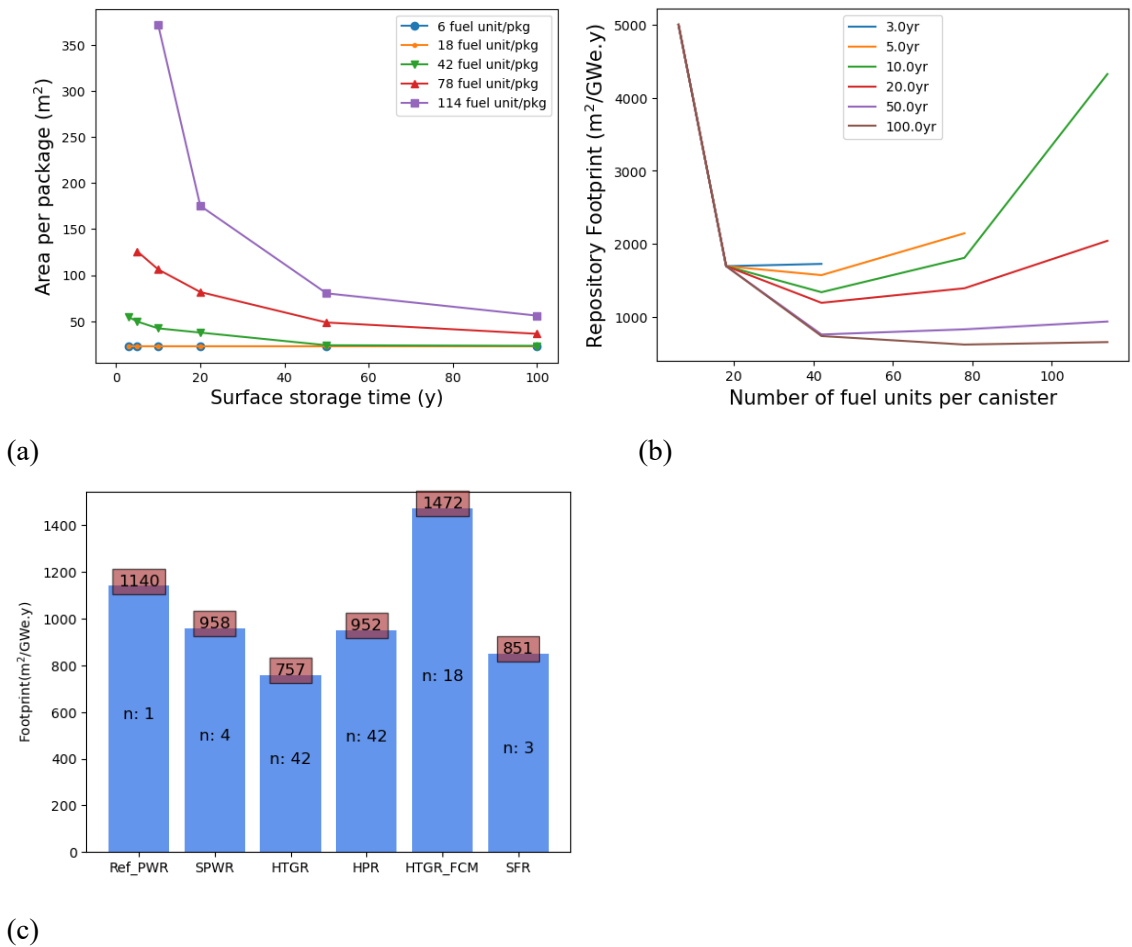

**Figure 4.** Repository footprint analysis results: (a) the required area per package for HTGR SNF in a clay repository as a function of storage time and package loading (i.e., the number of fuel elements in each package), (b) the repository footprint per GWe.y as a function of storage time and package loading, and (c) the minimum footprint per GWe.y equivalent SNF for each reactor after the storage of 50 years. In (c), the number of fuel elements in each package achieving this minimum is indicated in the middle of the bars (n).

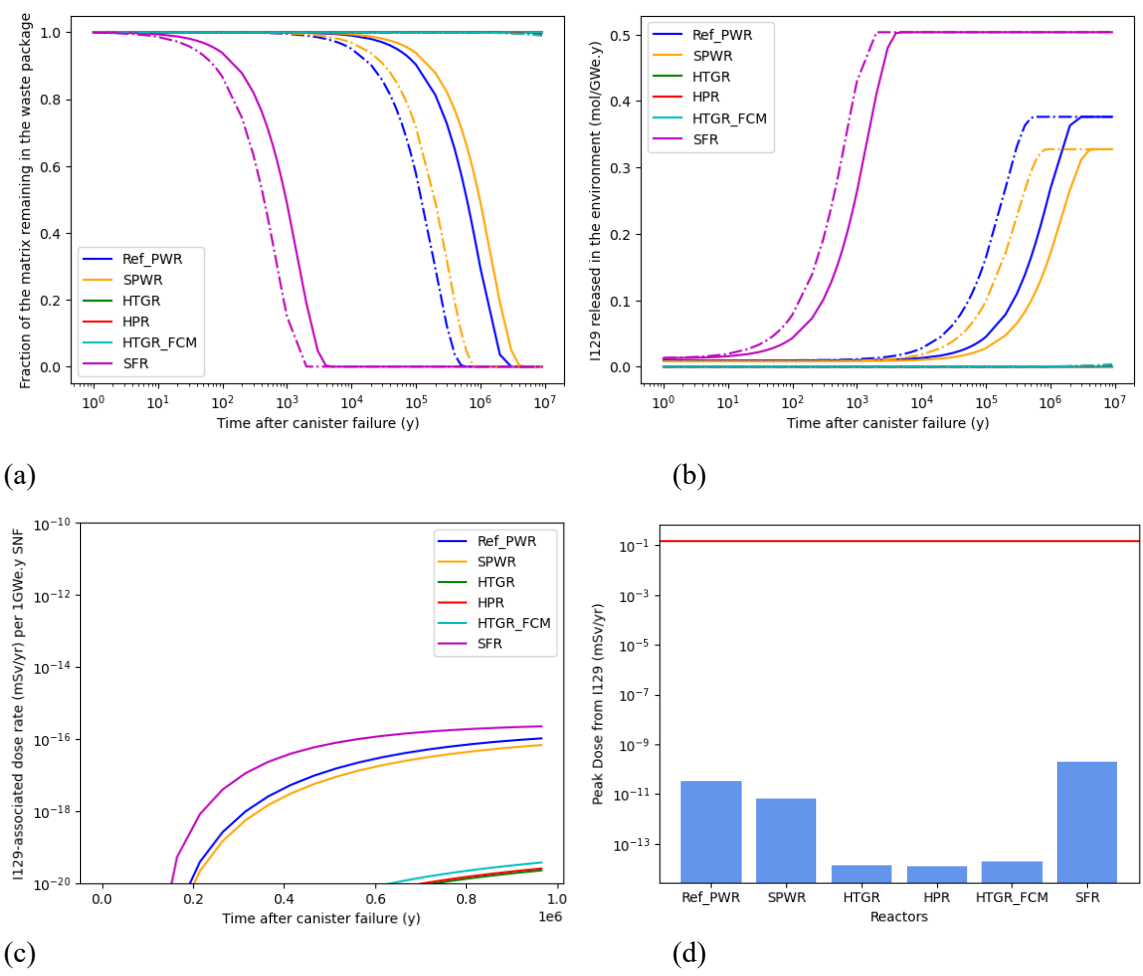

**Figure 5.** Repository PA results: (a) the fraction of the waste form matrix remaining in the waste package after contact with groundwater (i.e., canister failure), (b) the cumulative release of I-129 from the package to the geosphere, (c) the I-129-associated annual dose rates from the repository of the 1GW.y equivalent SNF, and (c) the peak annual dose rate from I-129 from the repository with 70,000 MTHM (metric ton of heavy metal). In (a) and (b), the dot-dashed lines correspond to the high degradation rate, while the plain line corresponds to the low degradation rate. In (d), the Yucca Mountain standard (0.15 mSv/yr) corresponds to the horizontal red line.

**Table 1.** SMR/MNR characteristics and technologies in the previous comparative studies. *Values calculated using the thermal efficiency. **Krall et al (2022) did not perform depletion calculations; the results are from MOX-based fuel (Kuwagaki et al., 2020).

| Paper | Reactor | Thermal Power, MWth | Electric Power, MWe | Thermal efficiency | Burnup, GWd/MT | Enrichment, % | Fuel | Neutron spectrum |
|---|---|---|---|---|---|---|---|---|
| Krall et al., 2022 | AP1000 Ref. PWR | 3400 | 1122* | 0.33 | 60 | 4.8 | UO2 | Thermal |
| | NuScale | 160 | 52.8* | 0.33 | 34 | 5 | UO2 | Thermal |
| | ISMR | 400 | 194* | 0.49 | 14 | 3 | Molten salt | Thermal |
| | SFR 4S30 | 30 | 9.9* | 0.33 | 34 | 19 | U-10Zr** | Fast |
| Kim 2022 | Ref. PWR | 3500* | 1175 | 0.34 | 50 | 4.5 | UO2 | Thermal |
| | VOYGR NuScale | 250* | 77 | 0.31 | 49.5 | 4.95 | UO2 | Thermal |
| | Natrium | 842* | 345.22 | 0.41 | 146 | 16.5 | U-Zr | Fast |
| | Xe-100 | 200* | 80 | 0.4 | 169 | 4.95 | TRISO | Thermal |
| Keto 2022 | Ref. PWR | 4300* | 1600 | 0.37 | 45 | 4.5 | UO2 | Thermal |
| | NuScale | 250* | 77 | 0.31 | 40 | 4.55 | UO2 | Thermal |
| Brown et al., 2017 | Ref. PWR | 3400 | 1122* | 0.33 | 30.5 | 4.95 | UO2 | Thermal |
| | NuScale | 400 | 132* | 0.33 | 30.5 | 4.95 | UO2 | Thermal |
| | NuScale | 160 | 52.8* | 0.33 | 32.4 | 4.95 | UO2 | Thermal |

**Table 2.** Selected SNF metrics studied for SMRs/MNRs. * Re-calculated to convert from per GW-thermal-year to per GW-electric-year.

| Paper | Reactor | Mass, t/GWe.y | Volume, m3/GWe.y | Activity 100 y, MCi/GWe.y | Activity 10,000 y, kCi/GWe.y | Decay heat 100y, kW/GWe.y | Radiotoxicity 10,000 y ×10$^8$ Sv/GWe.y |
|---|---|---|---|---|---|---|---|
| Krall et al., 2022 | AP1000 Ref. PWR | 18.4* | 6.06* | | | 11.7* | 1.18* |
| | NuScale | 32.5* | 15.5* | | | 9.1* | 1.39* |
| | ISMR | 53.8* | 22.7* | | | 5.6* | 1.37* |
| | 4S30 | 32.5* | 6.1* | | | 17.7* | 5.45* |
| Kim 2022 | Ref. PWR | 21.7 | 9.6 | 1.32 | 1.61 | 9.8 | 1.21 |
| | NuScale | 23.9 | 10.4 | 1.43 | 1.73 | 10.3 | 1.27 |
| | Natrium | 6.1 | 5.6 | 0.93 | 1.89 | 4.7 | 1.78 |
| | Xe1000 | 5.4 | 118.0 | 1.06 | 0.93 | 6.4 | 0.41 |
| Keto 2022 | Ref. PWR | | | | | 9.8 | |
| | NuScale | | | | | 9.7 | |
| Brown et al., 2016 | Ref. PWR | 22.1* | | 1.34* | 1.51* | | |
| | NuScale (400MWth) | 36.3* | | 1.40* | 1.69* | | |
| | NuScale (160 MWth) | 34.2* | | 1.35* | 1.69* | | |

**Table 3.** Selected SNF metrics studied for SMRs/MNRs. * Re-calculated to convert from per GW-thermal-year to per GW-electric-year.

|  | Fuel | Enrichment (%) | Thermal efficiency (%) | Volume-to-mass ratio | Burnup MWd/kgU | Reference |
|---|---|---|---|---|---|---|
| Ref. PWR | UO2 | 4.5 | 33 | 0.43 | 50 | Westinghouse, (2011) |
| SPWR | UO2 | 4.5 | 33 | 0.43 | 15 | Shirvan et al., (2023) |
| HTGR | UCO-TRISO | 15.5 | 40 | 21.8 | 80 | Shirvan et al., (2023) |
| HPR | UCO-TRISO | 19.75 | 33 | 21.8 | 80 | Shirvan et al., (2023) |
| FCM-HTGR | FCM-TRISO | 19.75 | 33 | 21.8 | 80 | Shirvan et al., (2023) |
| SFR | Metallic U-20Pu-10Zr | 20 | 40 | 0.912 | 120 | Kim (2020) |

**Supporting Information for**

Cross-disciplinary Reactor-to-Repository Framework for Evaluating Spent Nuclear Fuel from Advanced Reactors

Haruko M. Wainwright[1,2], Chloe Christiaen[3], Milos Atz[4], John Sebastian Tchakerian[1], Jiankai Yu[1], Gavin Keith Ridley[1], Koroush Shirvan[1]

[1] Department of Nuclear Science and Engineering, Massachusetts Institute of Technology

[2] Department of Civil and Environmental Engineering, Massachusetts Institute of Technology

[3] École Polytechnique

[4] Ultra Safe Nuclear Corporation

* Haruko M. Wainwright

**Email:** hmwainw@mit.edu

**Text S1. SNF Metrics**

The SNF metrics common across these studies (Brown et al., 2017; Keto et al., 2022; Kim et al., 2022; Krall et al., 2022) are primarily form the reactor physics simulations: SNF mass and volume, SNF radioactivity, decay heat, and radiotoxicity.

**SNF Mass (ton-HM/GWe.y):** The SNF mass (heavy metal equivalent) includes all heavy metals and fission products derived from the initial fuel materials (Wigeland, 2014). The SNF mass ($M_{SNF}$) can be obtained directly with the following equation:

$$M_{SNF}\left(\frac{t}{GWe.y}\right) = \frac{365 \text{ (d/y)}}{\text{Burnup (GWd/MTU)} * \text{Thermal efficiency}} \quad (S1)$$

This metric is relevant to evaluate SNF handling, storage, transportation, and final disposal. We would note that it only reflects the mass of heavy metal (uranium, actinides, and fission products) in SNF, and does not include other components of the core that are also stored in geological disposal, such as cladding materials for LWRs or graphite used as the fuel matrix for HTGR.

**SNF volumes (m³/GWe.year):** There are multiple approaches to quantify SNF volumes. We present two methodologies used in the previous studies for completeness.

In Krall et al. (2022), the entire volume of the active core is used to estimate the SNF volume to be disposed with the following equation:

$$\text{SNF Volume} \left(\frac{m^3}{GWth.year}\right) = \frac{\text{Volume of the active core }(m^3)}{\text{Thermal power (GWth)} * \text{residence time of the fuel (year)}} \quad (S2)$$

Taking the entire volume of the core induces not only the fuel matrix, claddings, and fuel unit materials (assembly or fuel block), but also the voids between the fuel elements. This approximation may lead to the overestimation of the SNF volume.

Kim et al. (2022) used the assembly or pebble volume-to-mass ratio (m³/MT) to compute the SNF volume in the following equation:

$$V_{SNF} = M_{SNF} * f_{mass}^{volume} \quad (S3)$$

where $V_{SNF}$ is the SNF volume in m³/GWe.year, $M_{SNF}$ is the SNF mass obtained in Equation (S1), and $f_{mass}^{volume}$ is the assembly or pebble volume-to-mass ratio in m³/t (volume of an assembly divided by the heavy metal loading per assembly). In this calculation, the volume is obtained at the assembly (or pebble) level, and therefore does not consider the voids between the assemblies (or fuel blocks).

**SNF Activity (Bq/GWe.year):** The radioactivity in SNF (per gram of initial uranium mass, gU) can be calculated according to the following equation:

$$\text{Activity} \left(\frac{Bq}{gU}\right) = \sum_{\text{Nuclide } i \text{ in SNF}} C_i \times \frac{\ln(2)}{t_{1/2,i}} \quad (S4)$$

where $C_i$ is the concentration of the nuclide $i$ in SNF (in atom/gU), and $t_{1/2\,i}$ is the half-life of the nuclide (in s). The normalized activity per electricity generated is then obtained with:

$$\text{Activity} \left(\frac{Bq}{GWe.year}\right) = \text{Activity} \left(\frac{Bq}{gU}\right) \times M_{SNF}\left(\frac{MT}{GWe.y}\right) \times 10^6 (MT/g) \quad (S5)$$

The SNF activity during the first 100 years—mainly governed by fission products—is relevant to radiation safety and others during discharge handling, packaging, storage, and transportation. The activity in the 1000-100,000 year interval is relevant to the repository performance.

**SNF Radiotoxicity (Sv/GWe.year):**

The radiotoxicity of SNF (Sievert) reflects the theoretical dose consequence of ingesting a particle of SNF, including all radionuclides present in SNF at a particular point in time. The radiotoxicity is defined as:

$$\text{Radiotoxicity} \left(\frac{\text{Sv}}{\text{gU}}\right) = \sum_{\text{nuclide } i \text{ in SNF}} A_i * DCF_i \qquad (S6)$$

where the $A_i$ in Bq/gU corresponds to the activity of each radionuclide, and $DCF_i$ is the Dose Conversion Factor for each nuclide $i$ (in Sv/Bq) from the International Commission on Radiological Protection (ICRP) publication (Clement, 2012). It reflects the health impact of ingesting a specific radionuclide by a member of the public. It is then converted to the radiotoxicity per energy generated in the same way as Equation (S5).

In the first several hundred years, fission products are dominant contributors to the radiotoxicity of the spent fuel. In the long term (between 10,000 and 100,000 years), transuranic isotopes are dominant contributors to radiotoxicity. The radiotoxicity at 100,000 years is a relevant metric to assess the long-term toxicity of SNF in the geological repository.

**SNF Decay Heat (W/GWe.year):**

Similar to the radioactivity, the overall decay heat in SNF is the sum of decay heat from all the radionuclides:

$$\text{Decay heat} \left(\frac{W}{\text{gU}}\right) = \sum_{\text{nuclide } i \text{ in SNF}} DH_i * C_i \qquad (S7)$$

where $DH_i$ (in W/atom) is the heat generated by a specified radionuclide due to several reactions, and $C_i$ is the concentration of the radionuclide in SNF (in atom/gU). It is then converted to the decay heat per electricity generated in the same calculation as Equation (S5).

The decay heat at 10–100 years is relevant for SNF handling, interim storage, transportation, and initial emplacement in the repository. Indeed, a surface storage time of 5 to 100 years is necessary before the final disposal to allow SNF heat to decrease.

**Text S2. Reactor Designs and Simulation Setups**

Reactors physics simulations determine the composition and the characteristics of SNF at various time points after the discharge. For each reactor, the simulations include core neutronics modelling, depletion analysis and the radioactive decay chain calculations in SNF. The key parameters are listed in Table 1, while the detail designs and computational setup are described in the following sections.

OpenMC is an open-source Monte Carlo neutron and photon transport simulation code recently developed at the Massachusetts Institute of Technology (Romano, 2013; Romano, 2015). It has the Python programming interface, which facilitates the connection to the repository models as well as the interactive workflow in Jupyter. We used the ENDF/B-VII.1 library for cross sections, depletion chain and decay chains, including the thermal and fast neutron spectrums (Chadwick et al., 2011). For the neutronics, this study uses 5,000 particles, with 200 active batches and 20 inactive batches. The CE/CM (constant extrapolation, constant midpoint) method is used for time integration.

The depletion calculation uses shorter depletion time steps at the beginning to accurately capture the build-up of xenon isotopes, and then longer depletion steps until the input burnup is reached. After the discharge from the reactor, shorter time steps are applied during the first hundred years to capture the decay of short-lived fission products. Then, longer time steps are applied up to 1,000,000 years after the disposal of SNF.

**Light Water Reactors (Reference PWR and Micro-PWR):**

$UO_2$-fueled SMRs/MMRs are based on reliable and available technologies as well as established supply chains. They are therefore expected to be deployed sooner than the other concepts (NuScale, 2020)

For the reference large LWR reactor and micro-PWR (MPWR), a single pin-cell was modelled and used to obtain SNF characteristics. The PWR pincell was directly adapted from the PWR pincell model available on the OpenMC tutorial. The parameters used are summarized in Table S1. The simulation was performed the height of a single fuel pitch, 1.26 cm, with periodic axial boundaries conditions to represent an infinitely tall core. The layout of the pin-cell is presented in Figure S1.

**High Temperature Gas Cooled Reactor with TRISO and FCM fuel:**

This study considers a generic HTGR design as a SMR/MMR based on the HALEU TRISO fuel, helium coolant, and graphite moderator. TRISO fuel particles are composed of a central kernel of uranium oxycarbide (UCO), which is surrounded by layers of porous carbon, pyrolytic carbon (PyC) and silicon carbide (SiC; Van den Akker and Ahn, 2013). In this study, we assume that the particles are embedded into a graphite matrix, and scattered randomly in fuel channels. Prismatic graphite blocks are then composed of fuel and coolant channels. The higher temperatures (700°C to 1000°C) lead to higher thermal-to-electricity efficiency. HTGR is known to have enhanced safety due to the low density of fissile material in the TRISO fuels, as well as the high temperature durability and heat-transfer properties of graphite (Atz, 2019).

The HTGR microreactor design was developed based on a scaling of the General Atomics' 350 MWth prismatic Modular HTGR (MHTGR) (General Atomics, 2012). A 3D homogeneous model of this core obtained from Kristina (2023) was scaled down and adapted to a microreactor design. In this design, the fuel pellets, which contains randomly dispersed TRISO particles embedded in a graphite matrix were homogenized while preserving the material volume, with the aim to reduce the computational time for time-dependent depletion calculation. We selected the homogeneous configuration with one region for TRISO fuel and layers and 2 graphite matrix regions, which was used in Kristina (2023).

In the microreactor design, several simplifications were made compared to the original model. The design of the core consists of an array of hexagonal fuel elements in a cylindrical arrangement, surrounded by a 40cm thick ring of reflector material (graphite) and the reactor vessel. The active core consists in a 3 rows annulus of prismatic fuel graphite blocks, contained within a 140 cm-diameter disk. The prismatic fuel block is made of a regular triangular array of 2 fuel holes, containing TRISO particles embedded in a graphite matrix, per one coolant hole filled with helium. The main specifications of the HTGR core are presented in Table S2. The simulations were performed on a 2D model. Reflective boundary axial conditions were set to represent an infinitely tall core.

Another HTGR design, using Fully Ceramic Microencapsulated (FCM) fuel, is considered in the study. The FCM fuel contains TRISO particles encased into a silicon carbide matrix (Duchnowski, 2022). This fuel has an advantage over the UCO-based TRISO because it may enhance safety during operating and accident conditions, with an increasing retention of fission products within the SiC matrix (USNC, 2021). In our study, we also assume that the fuel kernel material is replaced by uranium nitride (UN). Indeed, the SiC matrix results in a hardened neutron flux spectrum and requires a higher density fuel kernels such as UN (Lu, 2017).

The model described above was modified to be relevant with the HTGR using FCM fuel (called FCM-HTGR in our study). The active core diameter was increased to 220cm. The packing fraction, which corresponds to the mass of the TRISO particles divided by the mass of the fuel compact (mass of the matrix and mass of the particles), is increased to 0.60. The differences between the designs are summarized in Table S2. The core designs of the HTGR and FCM-HTGR models are presented respectively in Figure S2 and S3.

### Heat Pipe Reactor

Our HPR design is similar to that of HTGR with the HALEU TRISO fuel, with a power output ranging from 200 kW to 5 MW. The heat pipes are filled with liquid sodium to transport heat from the core to the heat exchanger. Heat pipes optimizes energy transfers and substitute the reactors coolant pumps and other auxiliary systems (Testoni, 2021).

The HPR model was based on the HTGR reactor model described above, with a similar 70 cm radius active core (Figure S4). Several adaptations were made to obtain a model for HPR: (a) heat pipes were modelled by surrounding the coolant channels with a 0.4 mm thick FeCrAl pipe (73wt% Fe, 20wt% Cr, 5wt% Al, 2wt% Mo), (b) the helium coolant was replaced by a sodium coolant with 50% void density, and (c) U235 enrichment was increased to 19.75% to compensate for the heat pipes absorption. A burnup of 80 MWd/kgHM and an efficiency of 33% were considered for this microreactor.

### Sodium Fast Reactor

Sodium cooled fast reactors (SFR) use liquid sodium as coolant, allowing higher operating temperature and increased thermal efficiency. Historically, SFRs have been considered in a closed fuel cycle with uranium and actinides recovery, as well as with partitioning and transmutation. However, we assume a once-through fuel cycle in this study for the comparison purposes. Their fuel is typically metallic Zr-alloy with stainless steel cladding and metallic fuel (Atz, 2019).

The SFR microreactor model was developed based on a scaling down of the 250 MWth Advanced Burner Test Reactor (Kim, 2020). The size, the components and the composition of the core are similar to the core design of the Toshiba-4S microreactor (Ueda et al., 2005).

The design of the core consists in an array of hexagonal elements in a prismatic arrangement (Figure S5). The 2 central prismatic rings are composed of fuel elements and are surrounded by 2 rings of hexagonal reflector elements and one ring of shield element. The outer most layer is the reactor pressure vessel. The fuel elements are composed of fuel pin cell disposed in a prismatic array and surrounded with sodium coolant. The fuel is a zirconium metal fuel U20Pu10Zr with a high enrichment in plutonium to increase the discharge burnup and to extend the reactor lifetime. The fuel is surrounded by a HT-9 cladding. Several simplifications were made compared to ABTR design: the active core is composed of one batch of fuel material and the control systems were removed from the core. In our 2D model, reflective axial boundary conditions were applied to be consistent with an infinitely tall core. The parameters of the SFR model are summarized in Table S3.

**Texts S3. Repository Footprint Model**

We follow the methodology developed by Atz (2019), which determines the canister spacing and repository footprint based on the decay heat generation and the repository conditions. The decay heat released from the waste packages could accelerate the degradation of EBS and the surrounding host rocks. To prevent such effects, a threshold temperature is typically set for the repository design, depending on the host rock and EBS properties. The approach in Atz et al. (2019) primarily focuses on the peak temperature at the surface of the canister located at the center of the repository, assuming it is higher than temperatures at any other location in the repository. The limit is 100 deg-C in the clay geologic setting, which has the most stringent temperature constraint due to the low heat conductivity of clay. This temperature limit is applied to prevent the thermally driven alterations of the bentonite buffer, which can increase rigidity, promote fractures, and decrease sorption as well as the mineralogical changes and thermally driven processes in the clay host rock.

In the footprint analysis, the canister designs are also important to determine the heat loading of each canister. The canisters usually consist of an inner canister, loaded with nuclear waste, and an overpack used to prevent corrosion. This work considers generic waste canisters and overpack designs (Table S4) specified in DOE (2012), in which the diameters were selected based on international accounts and previous conceptual design studies in the US. The TRISO-based SNF canisters are assumed to be loaded with prismatic fuel blocks containing TRISO particles. Even though the separation of the compacts from the graphite matrix is possible, this work assumes that the whole block is loaded in the canister. Although the canisters containing 78 and 114 fuel elements are not included in the previous studies, we explore increased loading, since the decay heat density of the TRISO-based SNF is relatively low. Finally, for SFR, we assume that the heavy metal loading is the same as the $UO_2$ fuel elements with the assemblies of the ABR-1000 reactor.

For the repository design, we considered the clay repository design reported in Atz (2019) and Hardin (2011). It is a horizontal disposal, with the initial drift spacing of 30 m. The containers with SNF are inserted into carbon steel overpacks and surrounded by bentonite buffer material. The emplacements drifts are then filled with clay buffer and back fill materials.

The footprint model is implemented in a Python library *npwy* developed by Atz (2019). Although the details are available in Atz (2019), we describe the model briefly for completeness. In the model, the temperature constraint is evaluated at the surface of a waste package located at the center of a square array of *N-by-N* waste packages (Figure 2a), where the heat contribution from adjacent packages is the highest. The drift spacing ($s_d$) and the package spacing within the drift ($s_p$) defines the minimum size of the repository. The package at the center of the array is modeled as a finite line source, while the remaining $N^2$-1 packages of the array are modeled as point sources.

The decay heat of each canister is given by $Q_{wf}(t)n_{wf}$, where $n_{wf}$ is the number of fuel units/elements in each canister, and $Q_{wf}(t)$ is the decay heat per fuel unit. With the uniform thermal conductivity κ and thermal diffusivity α, the changes in temperature at some distance *r* and time *t* due to the heat for a finite line source are given by the following analytical equation:

$$\Delta T_{fl}(r,t) = \int_0^t \frac{Q(t)}{8\pi k(t-\tau)L_{wp}} \cdot exp\left(-\frac{r^2}{4\alpha(t-\tau)}\right) \cdot erf\left(\frac{L}{4\sqrt{\alpha(t-\tau)}}\right) d\tau \qquad (S8)$$

The temperature increase due to the surrounding canisters (point sources) is given by:

$$\Delta T_{pt}(r,t) = \int_0^t \frac{Q(t)}{8\alpha^{\frac{1}{2}}\pi^{\frac{3}{2}}k(t-\tau)^{\frac{3}{2}}} \cdot exp\left(-\frac{r^2}{4\alpha(t-\tau)}\right) d\tau \qquad (S9)$$

For all times, the contributions from heat sources are superposed at the interface between the host rock and the last EBS layer. Each source is supposed to be surrounded by an infinite and homogeneous media which allows to superpose analytical solutions (Atz, 2019) for the impact of the surrounding $N^2$-1 packages on the center package.

The area per package is calculated through an optimization algorithm to finding the minimum area to satisfy the peak temperature constraint. The variable parameters include the drift spacing ($s_d$), the package space ($s_p$), and the size of the array *N*. The drift radius ($r_d$) and the package length ($L_{wp}$) are the fixed parameters of the repository design and the package loading. The algorithm minimizes the area per package, with respect to the constraints on the peak temperature and the minimum spacing requirements $s_p \geq L_{wp}$ and $s_d \geq 2r_d$. It uses the optimize.minimize function in the Python scipy package with the COBYLA (Constrained Optimization BY Linear Approximation) algorithm (Jones et al., 2001). Finally, the repository footprint is obtained by multiplying the area per package by the number of canisters that need to be placed in the repository.

In addition, we explore different surface cooling time before the disposal as well as the waste mass per package based on the previously studied package designs (DOE, 2012). We define the number of fuel elements (assemblies or fuel blocks) in each package, which yield the corresponding heavy metal equivalent and volume.

### Text S4. Repository Performance Assessment Model

In terms of the environmental impact of the waste disposal, the repository performance assessment (PA) quantifies the radionuclide transport through the geosphere and released to the biosphere (Ahn and Apted, 2012). After the canister failure, the waste form limits the release of radionuclides into groundwater. The main release modes are: (1) the instantaneous release mode in which noble gases (Kr and Xe) and other volatile fission products (such as I-129) are released instantaneously (Metz et al., 2012), and (2) the congruent release mode in which the radionuclides are released at the same rate as the fuel matrix degradation. Radionuclides are then transported in groundwater and then potentially come in contact with humans, mainly through groundwater pumping and its use for drinking and irrigation.

Our framework includes a simplified PA model estimating the potential dose risk from a generic clay repository filled with SNF. We focus on I-129, because it is the highest dose contributor in previous assessments (Van den Akker and Ahn, 2013; Mariner et al., 2015; Sassani, 2020) due to its high instantaneous release fraction, long half-life, unlimited solubility and lack of sorption. The model consists of three steps to quantify: (1) the I-129 inventory per canister, given the thermal constraint from the footprint analysis, (2) the release from the waste form in the congruent and instantaneous release modes, and (3) the transport in the geosphere.

In Step 1, we estimate the I-129 inventory per canister, based on the canister loading that minimize the repository footprint after a given surface storage time. The waste form mass within each canister is computed based on different waste forms. In addition, we calculate the number of canisters required for the 1GWe.y electricity production.

In Step 2, we assumed different waste-form dissolution rates and instantaneous release fraction, depending on the waste forms. The ranges of the dissolution rates (g/m$^2$.day) are [2.20E-03, 1.10E-02], [7.40E-08, 1.30E-06], and [1.57, 3.46] for the UO$_2$, graphite, and metallic fuel matrix, respectively. We assume that the instant release percentage for the UO$_2$ fuel is 2.5%, which is the value used in Mariner et al. (2015). We would note that this value is conservative, since it has been shown that for fuels with a burnup below 40 MWd/kgU the instant release fraction is under 1%. (Metz, 2012). For the TRISO and FCM fuels, the instant release fraction is supposed to be equal to $2.0 \times 10^{-4}$ (Petti et al., 2010), since the volatile fission products are encapsulated into the TRISO particles. Finally, the instant release fraction for the metallic fuels is assumed to be equal to that of the UO$_2$ fuel due to the lack of data.

The remaining I-129 fraction is released congruently with the fuel matrix. In this study, we use the model developed Ahn (2007) for the UO$_2$ and metallic fuel and Van der Akker and Ahn (2013) for the TRISO fuel to quantify the amount of I-129 released from a failed waste package in a geological repository. We would note that for TRISO particles, the major barrier preventing fuel degradation and radionuclide release is the graphite matrix of the fuel element, because the degradation rate of the graphite matrix is much lower than the one of PyC or SiC.

Here we describe the mathematical formulation of the congruent release for completeness, although the details were described in Ahn (2007) and Van der Akker and Ahn (2013). First, the geometrical transformation is applied for all matrix types to transform the fuel matrix into the sphere of the equivalent volume in a package. This approximation is then used to evaluate the evolution of the fraction of the matrix remaining in the canister. We assume a constant degradation rate.

For a sphere of radius $r$ (in m) and density $\rho$ (in kg/m$^3$) with the degrading at a rate $R$ (kg/m$^2$/s), the rate of change of the radius is given as:

$$\frac{dr}{dt} = -\frac{R}{\rho} \tag{S10}$$

$$r(t) = r_0 - \frac{R}{\rho}t \tag{S11}$$

where $r_o$ is the initial radius of the sphere calculated from the initial matrix mass contained in a canister. Assuming that the density of the matrix is uniform and does not vary in time, the fraction of the matrix remaining in the waste package is determined by the following equation:

$$\boldsymbol{f_{matrix}(t) = \left(1 - \frac{R}{\rho r_0}t\right)^3} \tag{S12}$$

Assuming that the concentration of I-129 is constant and uniform in the waste package, the mass of I-129 released out of the waste package congruently with the matrix is given as:

$$\boldsymbol{m_{I129}(t) = M_{I129}(1 - f_{matrix}) * (1 - IRF)} \tag{S13}$$

where $m_{I129}(t)$ is the mass of $I^{129}$ released in the environment after time $t$, $M_{I129}(t)$ is the mass of I-129 initially contained within the canister, and IRF is the instant release fraction.

In Step 3, we use a groundwater flow and contaminant transport simulator PFLOTRAN (Hammond et al., 2014; Lichtner et al., 2015), which has been extensively used in the SNF repository PA in the US (Mariner et al., 2015). Our conceptual model (Figure 2b) is a simplified version of the generic clay model described in Stein et al. (2018), such that only the clay host rock and limestone aquifer below are considered. The main flow parameters are included in Figure 2b. It does not include a bentonite buffer around each canister, because the key parameters, such as permeability, are comparable between the host rock and the buffer. Our grid spacing is 20 m by 20 m horizontally, and 10 m vertically. We assume that I-129 is a non-reactive species, which is consistent with previous studies (e.g., Mariner et al., 2015).

Because the repository footprint model shows that the 1GWe.y-equivalent footprint can fit within a few grid blocks, the source term is defined at one repository block based on the release rate from Step 2. Since the I-129 is not solubility limited, the concentration linearly increases with the number of canisters if they are arranged parallel to the groundwater flow (Kawasaki et al., 2005). If they are arranged perpendicular to the flow, the concentration does not depend on the number of packages. Since our focus is to compare SNF from different reactors, we consider that it is appropriate to consider the 1GWe.y-equivalent footprint without considering the repository configurations. As the performance metric, we consider the peak I-129 concentration in the well located 10 km from the repository, and associated annual dose rates. We convert the concentrations to the annual dose rates based on the biosphere dose conversion factor (for example, rem/yr divided by pCi/L) used for the Yucca Mountain (YM) assessment (Swift et al., 2008). We would note that the YM assessment assumed a well 18 km away from the repository as well as the current population and water use. We assume that the distance of 10 km is compatible to this assessment. To compare the peak annual dose rates with the YM standard, we scale the dose rates from per GWe.y to the YM capacity of 70,000 metric tons of heavy metal.

## SI References

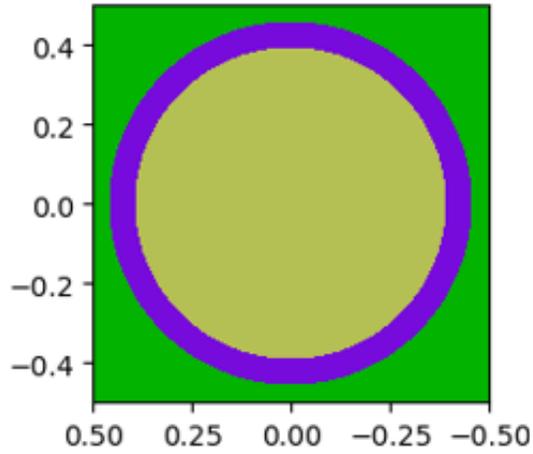

**Fig. S1. PWR Pincell Model on OpenMC. The light green region is the UO2 fuel, the purple region is the zircaloy cladding, and the green outer region is the borated water.**

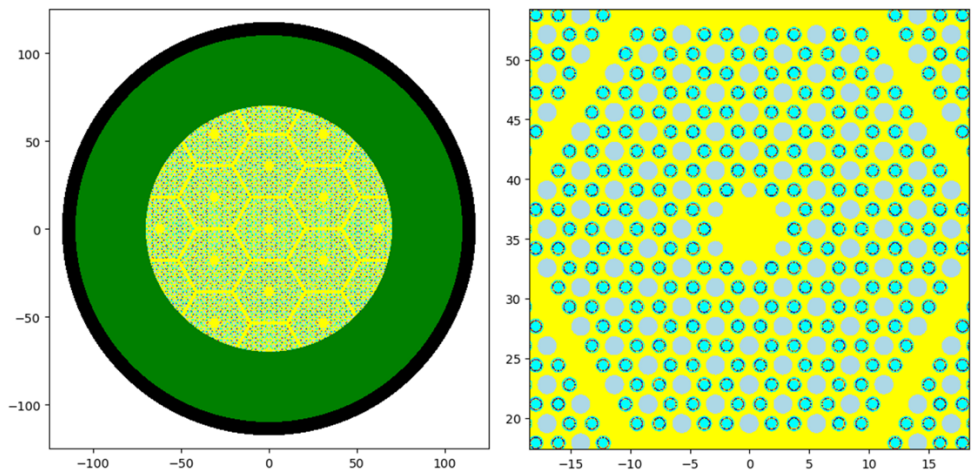

**Fig. S2. 2D view of OpenMC Model for HTGR microreactor. On the left, a view of the full core model is presented. On the right, a zoom is made on a prismatic fuel block.** *Dark: Vessel, Green: Graphite reflector, Yellow: Graphite, Light blue: coolant hole, Cyan: Graphite matrix, Purple: Homogeneous TRISO*

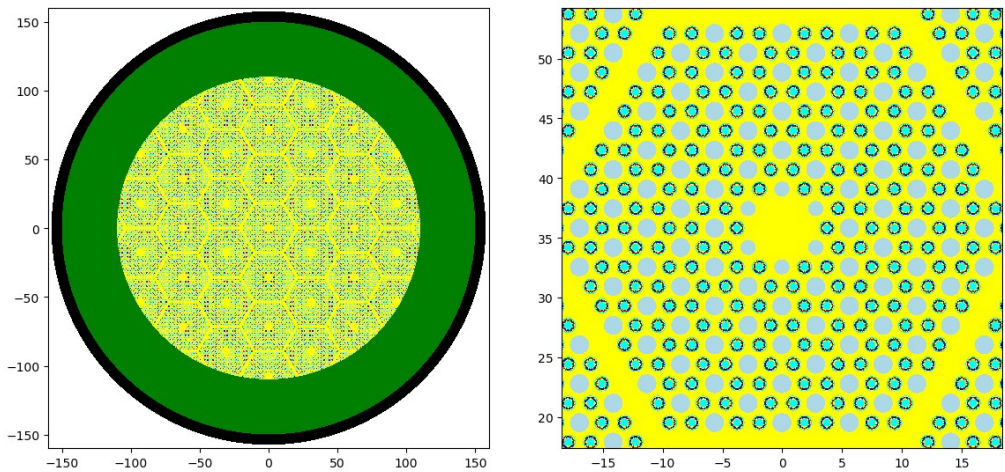

**Fig. S3. 2D view of OpenMC Model for the FCM-HTGR. On the left, a view of the full core model is presented. On the right, a zoom is made on a prismatic fuel block.** *Dark: Vessel, Green: Graphite reflector, Yellow: Graphite, Light blue:* coolant hole, Cyan: SiC matrix, Purple: Homogeneous TRISO.

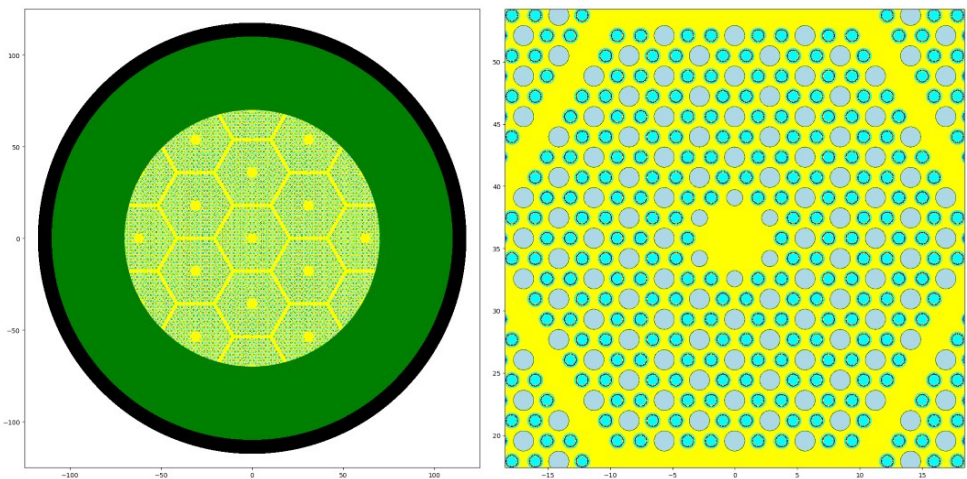

**Fig. S4. 2D view of OpenMC Model for Heat-Pipe microreactor. On the left, a view of the full core model is presented. On the right, a zoom is made on a prismatic fuel block. Dark: Vessel, Green: Graphite reflector, Yellow: Graphite, Light blue: Coolant hole with sodium, Cyan: Graphite matrix, Purple: Homogeneous TRISO, Brown: Coolant Pipe.**

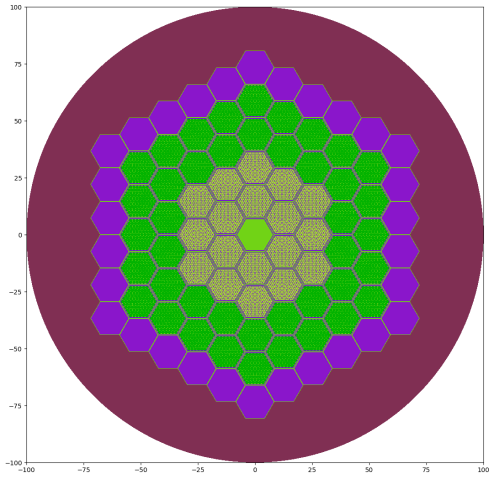

**Fig. S5. SFR core model designed on OpenMC. From the inner most to the outer most ring, the components are: Empty cell filled with coolant, fuel assembly, reflector, shield, vessel.**

**Table S1: Reference PWR and PWR microreactor input parameters and geometry on OpenMC.**

| Parameters | Reference PWR (Westinghouse, 2011) | Small PWR (Shilvan et al., 2023) | Unit |
|---|---|---|---|
| U-235 Enrichment | 4.5 | 4.5 | % |
| UO2 Fuel density | 10.5 | 10.5 | g/cm3 |
| Linear power | 190 | 130 | W/cm |
| Fuel outer radius | 0.40947 | 0.40947 | cm |
| Cladding outer radius | 0.45720 | 0.45720 | cm |
| Cladding material | Zircaloy | Zircaloy | |
| Coolant material | Borated water | Borated water | |
| Burnup | 50 | 15 | MWd/kgHM |
| Efficiency | 33 | 33 | % |

**Table S2: HTGR and FCM-HTGR reference configuration and parameters.**

| Parameter | HTGR (Shilvan et al., 2023) | FCM-HTGR | Unit |
|---|---|---|---|
| Burnup | 80 | 80 | MWd/kgHM |
| Efficiency | 40 | 33 | % |
| Power density | 5 | 1.25 | kW/L |
| Linear power | 77.0 | 47.5 | kW/cm |
| Number of fuel blocks | 18 | 26 | |
| Number of fuel holes per fuel blocks | 216 | 216 | |
| Coolant holes per fuel blocks | 108 | 108 | |
| Active core diameter | 140 | 220 | Cm |
| Fissile material | $UC_{0.5}O_{1.5}$ | UN | |
| U235 initial enrichment | 15.5 | 19.75 | wt% |
| TRISO coating layers materials | Kernel/Buffer/ iPyC/SiC/oPyC | Kernel/Buffer/ iPyC/SiC/oPyC | |
| Homogeneous TRISO coating layers radius | 212.5/312.5/347.5/382.5/422.5 | 364.3/535.7/ 595.7/ 655.7/ 724.3 | μm |
| TRISO coating layer densities | 10.5/1.0/1.9/3.2/1.9 | 14.3/1.0/1.9/3.2/1.9 | g/cm3 |
| TRISO packing fraction | 0.35 | 0.6 | |
| Fuel compact radius/with gap | 0.6225/0.635 | 0.6225/0.635 | cm |
| Fuel compact matrix (density) | Graphite (1.7) | SiC (3.2) | g/cm3 |
| Primary coolant | Helium | Helium | |
| Coolant hole radius | 0.795 | 0.795 | cm |
| Holes pitch | 1.8796 | 1.8796 | cm |
| Fuel block edge length | 20.6756 | 20.6756 | cm |
| Reflector thickness | 40.0 | 40.0 | cm |
| Fuel temperature | 900 | 900 | K |
| Coolant temperature | 600 | 600 | K |

| Operating pressure | 6.39 | 3.00 | MPa |

**Table S3: Main specifications of SFR core model**

| Parameter | SFR microreactor | Unit |
|---|---|---|
| Burnup | 120 | MWd/kgHM |
| Efficiency | 40 | % |
| Power density | 68.8 | kW/l |
| Linear power | 250 | kW/cm |
| Coolant | Sodium | |
| Number of fuel assembly | 18 | |
| Assembly edge lengh | 14.6850 | cm |
| Number of pin-cell per assembly | 217 | |
| Fuel pin pitch | 0.9080 | cm |
| Fuel pin diameter | 0.8000 | Cm |
| Fissile material (density) | U20Pu10Zr (15.8) | g/cm3 |
| Cladding material (density) | HT-9 (6.55) | g/cm3 |
| Cladding thickness | 0.0520 | cm |
| Reflector material (density) | HT-9 (6.55) | g/cm3 |
| Shield material (density) | HT-9 and B4C (4.14) | g/cm3 |
| Fuel temperature | 855 | K |
| Coolant temperature | 700 | K |
| Operating pressure | 6.39 | MPa |

**Table S4: Spent fuel package dimensions and loading.**

| Reactor | Number of fuel elements (assemblies/ fuel blocks per canister) | Dimensions (m) (Diameter x Length) Thickness = 0.125m | Heavy metal mass (kg/assembly) | Total Heavy Metal (kg/canister) |
|---|---|---|---|---|
| PWR | 1 | 0.41 x 5.0 | 460 | 460 |
|  | 2 | 0.82 x 5.0 |  | 920 |
|  | 3 | 0.82 x 5.0 |  | 1380 |
|  | 4 | 0.82 x 5.0 |  | 1840 |
|  | 12 | 1.29 x 5.0 |  | 5520 |
|  | 21 | 1.60 x 5.0 |  | 9660 |
| HTGR | 6 | 0.6725 x 4.8875 | 8.6 | 51.6 |
|  | 18 | 1.0548 x 4.8875 |  | 154.8 |
|  | 42 | 1.3141 x 4.8875 |  | 361.2 |
|  | 78 | 1.9818 x 4.8875 |  | 670.8 |
|  | 114 | 1.9818 x 4.8875 |  | 980.4 |
| FCM-HTGR | 6 | 0.6725 x 4.8875 | 21.2 | 127.2 |
|  | 18 | 1.0548 x 4.8875 |  | 381.6 |
|  | 42 | 1.3141 x 4.8875 |  | 890.4 |
|  | 78 | 1.9818 x 4.8875 |  | 1654 |
|  | 114 | 1.9818 x 4.8875 |  | 2417 |
| SFR | 1 | 0.46 x 5.0 | 97.7 | 97.7 |
|  | 3 | 0.66 x 5.0 |  | 293.1 |
|  | 7 | 0.77 x 5.0 |  | 683.9 |
|  | 19 | 1.11 x 5.0 |  | 1856.3 |